# A compact and efficient three-dimensional microfluidic mixer

*Wenbo Li, Wei Chu\*, Peng Wang, Jia Qi, Zhe Wang, Jintian Lin, Min Wang and Ya Cheng\**

Wenbo Li, Dr. Wei Chu, Peng Wang, Jia Qi, Zhe Wang, Dr. Jintian Lin, Prof. Ya Cheng
State Key Laboratory of High Field Laser Physics
Shanghai Institute of Optics and Fine Mechanics
Chinese Academy of Sciences
Shanghai 201800, People's Republic of China
E-mail: ya.cheng@siom.ac.cn
Wenbo Li, Zhe Wang, Prof. Ya Cheng
School of Physical Science and Technology
ShanghaiTech University
Shanghai 200031, People's Republic of China
Wenbo Li, Dr. Wei Chu, Peng Wang, Jia Qi, Zhe Wang, Dr. Jintian Lin, Prof. Ya Cheng
University of Chinese Academy of Sciences
Beijing 100049, People's Republic of China
E-mail: chuwei0818@qq.com
Peng Wang
School of Physics Science and Engineering
Tongji University
Shanghai 200092, China
Dr. Min Wang, Prof. Ya Cheng
XXL - The Extreme Optoelectromechanics Laboratory
School of Physics and Electronic Science
East China Normal University
Shanghai 200241, China
Dr. Min Wang, Prof. Ya Cheng
State Key Laboratory of Precision Spectroscopy
East China Normal University
Shanghai 200062, People's Republic of China
Prof. Ya Cheng
Collaborative Innovation Center of Extreme Optics
Shanxi University
Taiyuan, Shanxi 030006, People's Republic of China

**Abstract**

Microfluidic mixing is a fundamental functionality in most lab on a chip (LOC) systems, whereas realization of efficient mixing is challenging in microfluidic channels due to the small Reynolds numbers. Here, we design and fabricate a compact three-dimensional (3D) micromixer to enable efficient mixing at various flow rates. The performance of the fabricated micromixer was examined using different combinations of liquid samples, including blue and red inks as well as purified water and a microsphere suspension. The extreme flexibility in fabricating microfluidic structures of arbitrary 3D geometries using femtosecond laser micromachining allows us to tackle the major disadvantageous effects for optimizing the mixing efficiency.

**Introduction**

Mixing plays a key role in chemical reaction. Microfluidic technology provides an effective means to the realization of highly efficient mixing of liquids by manipulating micro- and nanoscale fluids in sophisticated manners. [1-6] Various geometries have been incorporated into the microfluidic channels for promoting mixing efficiency including T-shaped microchannel, H-shaped micromixer, Grooved micromixer, etc.[7-12] In particular, it has been demonstrated that a three-dimensional (3D) passive micromixer, which is designed based on the Baker's transformation concept, can enable fast and efficient mixing even in the low-Reynolds-number condition. [13] It should be mentioned that the 3D micromixer is fabricated using femtosecond laser internal processing of glass, [14-18] which has been proved to be a straightforward approach for fabrication of 3D microfluidic structures and integrated optofluidic devices. [19-22]

For the sake of miniaturization and high-density integration, there is always interest to downsize the micromixer by promoting the mixing efficiency and improving the design strategy. Here, we demonstrate a compact and efficient 3D micromixer based on the Baker's transformation concept. Our micromixer is accommodated in a 1.6 cm-long channel of a rectangular cross-sectional size of 1 mm × 1 mm. In particular, the boundary effect in the microfluidic channels, which can cause a substantial degradation in the mixing process, is circumvented by periodic exchanging the microstreams on the left and right-handed sides in the microfluidic channel. We show excellent mixing performance of the fabricated micromixer by examining the mixing efficiencies of two kinds of ink solutions of different colors.

**Device design and simulations of mixing process**

**Figure 1** schematically shows the design principle of the 3D microfluidic mixer. The device is composed of a string of mixing units categorized into segments S1 and S2, as shown in Figure 1(a). The first segment (S1), as illustrated in Figure 1(b), is designed to increase the number of microstreams in the microchannel from *N* to *2N* as enabled by the Baker's transformation. The working principle of S1 is described as follow (i.e, using the first S1 unit as an example). Two different microstreams (i.e., illustrated in yellow and blue in Fig. 1) to undertake the mixing process are simultaneously sent into the first S1 unit from the two inlets of micromixer. The two microstreams are separated from each other as the right (in blue) and left (in yellow) microstreams. At the entrance of S1, the two microstreams are divided into the upper and lower streams, i.e., one going upwards and the other going downwards in two 3D mirochannels. The 3D channels then transform the upper and lower microstreams to the right and left channels at the exit of S1. As a result, the two microstreams at the inlet will be split into four microstreams consisting of alternatively arranged blue and yellow microstreams as indicated in Figure 1(b). Repeating of the process in S1 can result in rapid increase of the number of microstreams in the microchannel as a function of $N \sim 2^n$, where *n* is the number of S1 segments used in the construction of the micromixer.

The second segment S2 as illustrated in Figure 1(c) consists of two twisted channels which relocates the microstreams in the left-handed region to the right-handed region and vice versa. Therefore, the functionality of S2 is not to increase the number of microstreams but to move the microstreams which are initially in close contact to the two sidewalls of the microchannel at the entrance of S2 to the center of the microchannel at the exit of S2. The fluidic dynamics of microstreams can be affected by the sidewall of microchannel, resulting in lower mixing efficiencies for the microstreams on the two sides than that in the center area. By balancing the mixing efficiencies of the microstreams near the two sides and that in the center area with S2, the overall mixing efficiency can be effectively improved as compared with a micromixer composed of only S1. We will numerically prove this below.

**Figure 2** presents the simulated mixing performances in three microfluidic mixers. The numerical simulations were carried out by solving the microfluidic incompressible Navier-Stokes and convection diffusion equations using a finite element analysis software (COMSOL Multiphysics 5.4). In our simulation, the model micromixer in Figure 2(a) is simply a 1D straight channel of a cross sectional size of 1 mm×1 mm and a length of 1 cm. The concentration, flow rate and diffusion coefficient of the two kinds of liquids were set as 1 mol L$^{-1}$, 3 mm s$^{-1}$ and 4.5×10$^{-9}$ m$^2$ s$^{-1}$, respectively. To evaluate the mixing efficiency, we introduce the relative concentration variance ($R_{CV}$), which can be expressed as:

$$R_{CV} = \frac{R_{out}}{R_{in}}, \quad (1)$$

where $R_{in}$ and $R_{out}$ are the integrals of the concentration differences in the cross sections of the entrance and the exit of the mixer respectively, which can be further

written as:

$$R_{in} = \iint (c - c_0)^2 \, dA / \iint dA, \tag{2}$$

$$R_{out} = \iint (c - c_0)^2 \, dA / \iint dA. \tag{3}$$

Here $c$ is the localized concentration in the cross sectional plane, and $c_0$ is the concentration foe two fluids which are thoroughly mixed, and $dA$ is the differential area in the cross section plane. For ideally (i.e., thoroughly) mixed fluids, we have the relative concentration variance $R_{CV}=0$. Figure 2(a) shows that the mixing process in the 1D microchannel only occurs at the interface of two fluids owing to the characteristic laminar flow nature in a microfluidic channel and the limited diffusion coefficient of the fluids. The calculated $R_{CV}$ was 0.52 in Figure 2(a).

3D micromixers can significantly improve the mixing efficiency due to the mechanisms mentioned above. First, we examine the mixing performance in a model mixer consisting of only S1 (i.e., a straight string of six segments S1), as shown in Figure 2(b). In particular, the dimensions of S1 are indicated in the inset. It can be seen that after each pass of the microstreams through the segment S1, the fluids in the micromixer appear more uniformly distributed which indicates occurrence of efficient mixing. For the two fluids illustrated in red and blue, mixing leads to generation of the mixture in green. It can also be seen from Figure 2(b) that mixing initiates from the interface of the two fluids and gradually spreads away from the interface. Meanwhile, the microstreams near the two sides show a low mixing efficiency which remain insufficiently mixed at the exit of each segment of S1. This is because that the microfluids undergo a much higher flow rate in the center area than in the areas near the two sidewalls of the microchannel. The higher flow rate in turn gives rise to a higher mixing efficiency preferentially in the center area. The calculated $R_{CV}$ was 0.035 in Figure 2(b), which is more than one order of magnitude higher than the calculated $R_{CV}$ in Figure 2(a).

To solve the problem revealed by the result in Figure 2(b), we design a micromixer consisting of four segments of S1 and two segments of S2 which are arranged in an order of S1-S1-S2-S1-S1-S2, as schematically illustrated in Figure 2(c). The dimensions of S1 are the same as that in Figure. 2(b), thus only the dimensions of S2 are indicated in the inset of Figure. 2(c). The $R_{CV}$ in Figure 2(c) is calculated to be 0.0035, which is further reduced by one order of magnitude as compared with the result in Figure 2(b). It is noteworthy that the two 3D micromixers in Figure 2(b) and 2(c) have a same channel length and a same cross-sectional area. Thus, the throughputs of the two micromixers are essentially the same which is critical for evaluating and comparing the mixing efficiencies in the two mixers.

**Fabrication**

A 3D femtosecond laser micromachining system as schematically illustrated in Figure 3(a) was used for fabricating the 3D micromixers. The femtosecond laser pulses (1130 nm, up to 400 µJ, 270 fs) were provided by a commercial femtosecond laser source (Pharos, Light Conversion Ltd.). The duration of the laser pulse can be tuned from 270 fs to 15 ps by adjusting the distance between the gratings in compressor. After passing through an attenuator and a beam expanding system, the laser pulses were then focused into the fused silica glass using an objective lens (Olympus MPLFLN, 20×, NA = 0.45). A motion stage (ANT130-110-L-ZS, Aerotech Inc.) was used to translate the objective lens along Z direction to control the depth of the focus position, and the fused silica glass sample was mounted on an XY motion stage (ABL15020WB and ABL15020, Aerotech Inc.) and smoothly translated at a positioning precision of 100 nm. Both the translation stages were controlled using the high-performance motion controller (A3200, Aerotech Inc.).

In our fabrication, the repetition rate of the laser was set to 100 kHz, and the laser pulses were set as 4 ps. [23-24] We scanned the laser focal spot along the pre-designed paths layer by layer with a layer spacing of 10 μm to produce the 3D micromixer. The scan process was performed from the bottom to the top of the glass and the scan speed was fixed at 10 mm s$^{-1}$. After the laser irradiation, the glass samples were immersed in a solution of potassium hydroxide (KOH) with a concentration of 10 mol L$^{-1}$ for selectively removing the glass material exposed to the irradiation of the laser pulses. In total, it took ~24 hours to produce the micromixer. The sample was cleaned by a plasma cleaner and then mildly annealed at 100 °C for ~2 hrs. The procedures of the laser fabrication were schematically illustrated in Figure. 3(b).

**Figure 4** (a) presents the computer design of the 3D micromixer consisting of four segments of S1 and two segments of S2. Figure 4(b) and 4(c) show the top-view and side-view optical micrographs of the fabricated micromixer, respectively. The consistency between the fabricated and designed devices indicates a high fabrication resolution offered by the femtosecond laser 3D micromachining. There is no microcracks observed under the optical microscope. The complex 3D structures of micromixer as indicated by the two frames F1 and F2 in Figure 4(b) and (c). The detailed features of F1 are further clarified with the top view and side view micrographs in respective Figure 4(d) and (e), and the detailed features of F2 are correspondingly shown in Figure 4(f) and (g). The 3D features of complex geometries confirm the extreme flexibility of femtosecond laser micromachining of 3D microdevices in glass with unprecedented resolutions.

**Results and discussion**

We carried out microfluidic mixing experiments with the fabricated 3D micromixer. We simultaneously injected two kinds of ink solutions of different colors (yellow and blue) into the micromixer from the two inlets. We tested the mixing efficiency at different flow rates. The flow rate was controlled using a syringe pump. The mixing process was monitored using a home-made microscope. First, the injection rate was set to be 3 ml min$^{-1}$, which corresponds to a flow rate of 10 cm s$^{-1}$ and a Reynolds number of 100 in the micromixer. Figure 5(a) and 5(b) compare the mixing efficiencies in the straight microchannel and the fabricated 3D micromixer, respectively. The two devices have a same length and a same cross-sectional size. The mixing behaviors in Figure 5(a) and (b) agree well with the simulation results in Figure 2(a) and (c), respectively. Namely, the two fluids in yellow and blue were well separated throughout the 1D straight channel, indicating that at a Reynolds number of 100, the fluidic dynamics is dominated by laminar flow. The mixing highly relies on molecular diffusion at the interface, which is known to be a relatively slow process. In contrast, highly efficient mixing was confirmed in the 3D micromixer as we can expect from the model analysis and numerical simulation in Figure 2(c). It is striking that the experimental result in Figure 5(b) shows that mixing is inefficient near the two sidewalls of the microchannel. Thus, exchange of the microstreams in the center and side areas with S2 becomes necessary as evidenced in Figure 5(b).

To examine the mixing performance at various flow rates, we increased the injection rate to 10 ml min$^{-1}$, resulting in a flow rate of 30 cm s$^{-1}$ and a Reynolds number of 330. Figure 5(c) and 5(d) show that at the increased flow rate, the mixing behaviors in the respective straight and 3D micromixers maintain the same characteristics as observed at the low flow rates in Figure 5(a) and 5(b). In both cases, efficient mixing has been achieved in the 3D micromixer. However, if one carefully examines the mixing behaviors in Figure 5(a) and (c), it can be seen that the mixing efficiency slightly increases in the 1D straight channel with the increase of flow rate. This can be attributed to the generation of micro-turbulence which is more likely to occur at large Reynolds numbers.

**Conclusion**

To conclude, we have designed and fabricated a 3D microfluidic mixer using femtosecond laser micromachining. Our experiments show that the device can realize efficient microfluidic mixing for a wide range of Reynolds numbers. It is discovered, both experimentally and numerically, that the mixing efficiency can be significantly improved by periodically switching the microstreams in the middle and side areas in the microfluidic channel, as the diffusion near the sidewalls of the microchannel is inefficient. The compact and efficient 3D micromixer can be used in applications ranging from chemical/biological analysis and microfluidic synthesis of materials to fine chemistry microreaction.

**Acknowledgement**
The work is supported by Key Project of the Shanghai Science and Technology Committee (Nos. 18DZ1112700).

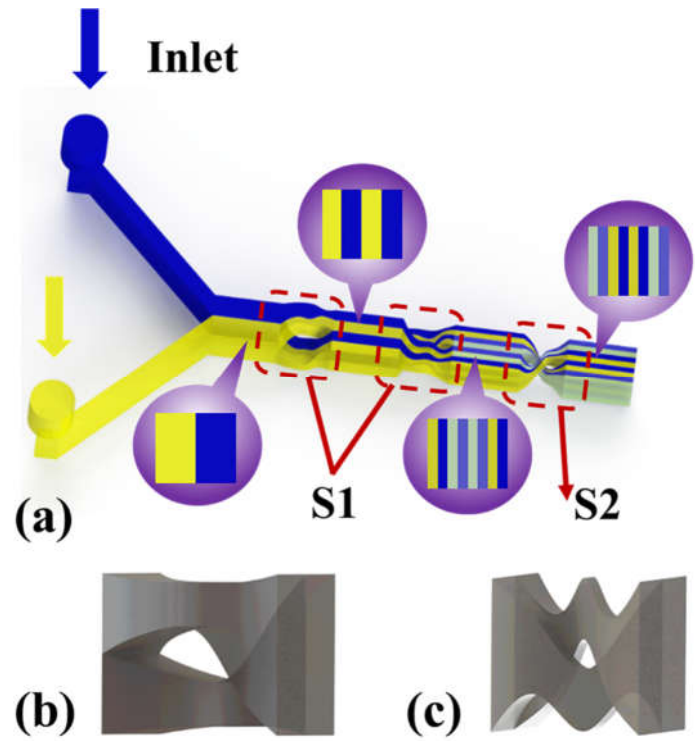

**Figure 1** (a) Schematic illustration of the design of 3D micromixer, and close-up illustrations of (b) S1 and (c) S2. Inset in (a): the cross-sectional distributions of the microstreams going through a string of S1 and S2.

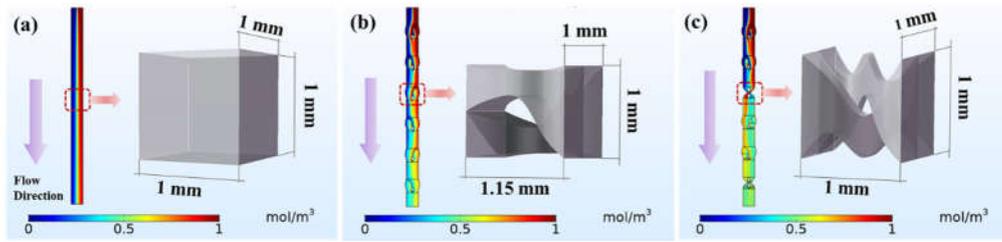

**Figure 2** Numerical simulations of mixing performances in (a) a 1D microchannel of square cross section; (b) a micromixer consisting of six segments of S1; and (c) a micromixer consisting of four segments of S1 and two segments of S2. All the micromixers have a same total length and a same cross sectional area.

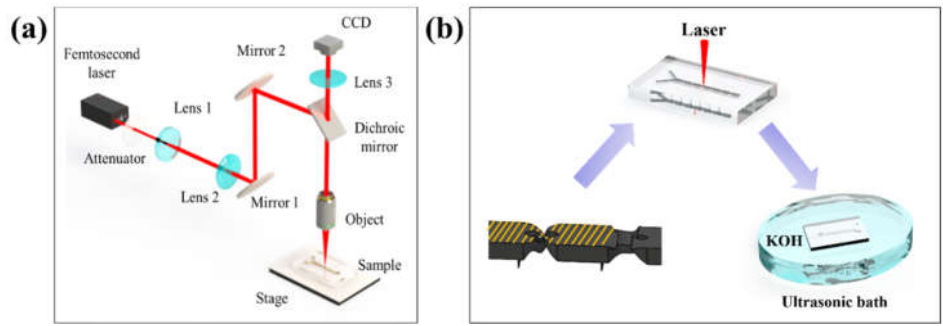

**Figure 3** (a) Schematic illustration of the experimental setup of 3D femtosecond laser machining system. (b) The major steps for the micromixer device fabrication: digitalization of the 3D model (left-handed panel), scan of the laser beam along the pre-designed paths to selectively modify glass (middle panel), and removal of the irradiated materials with the chemical wet etching (right-handed panel).

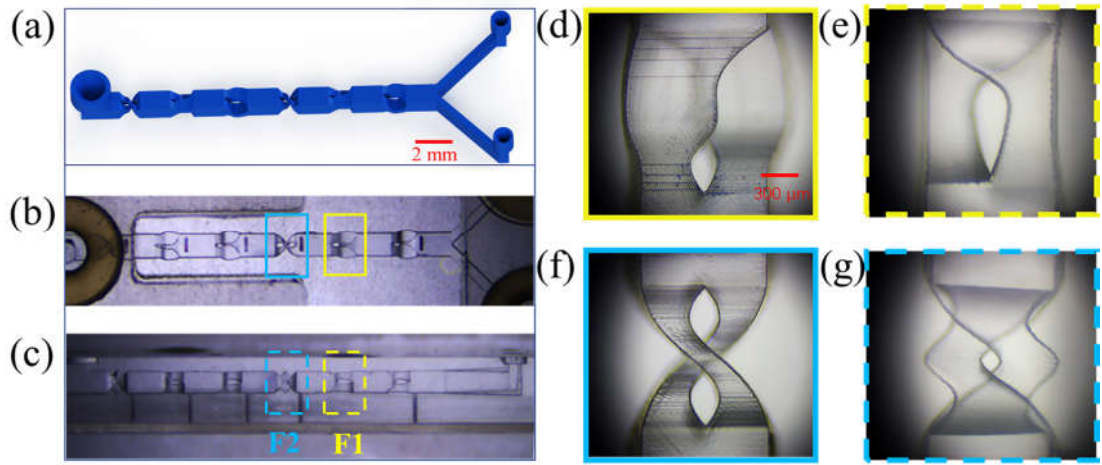

**Figure 4** Fabricated micromixer device in fused silica. (a) Illustraton of the model structure. (b) Top and (c) side view of the fabricated micromixer. The detailed features of F1 are shown in the (d) top view and (e) side view micrographs, and the detailed features of F2 are shown in the (f) top view and (g) side view micrographs. Scale bars: 2 mm in (a)-(c) and 300 μm in (d)-(g).

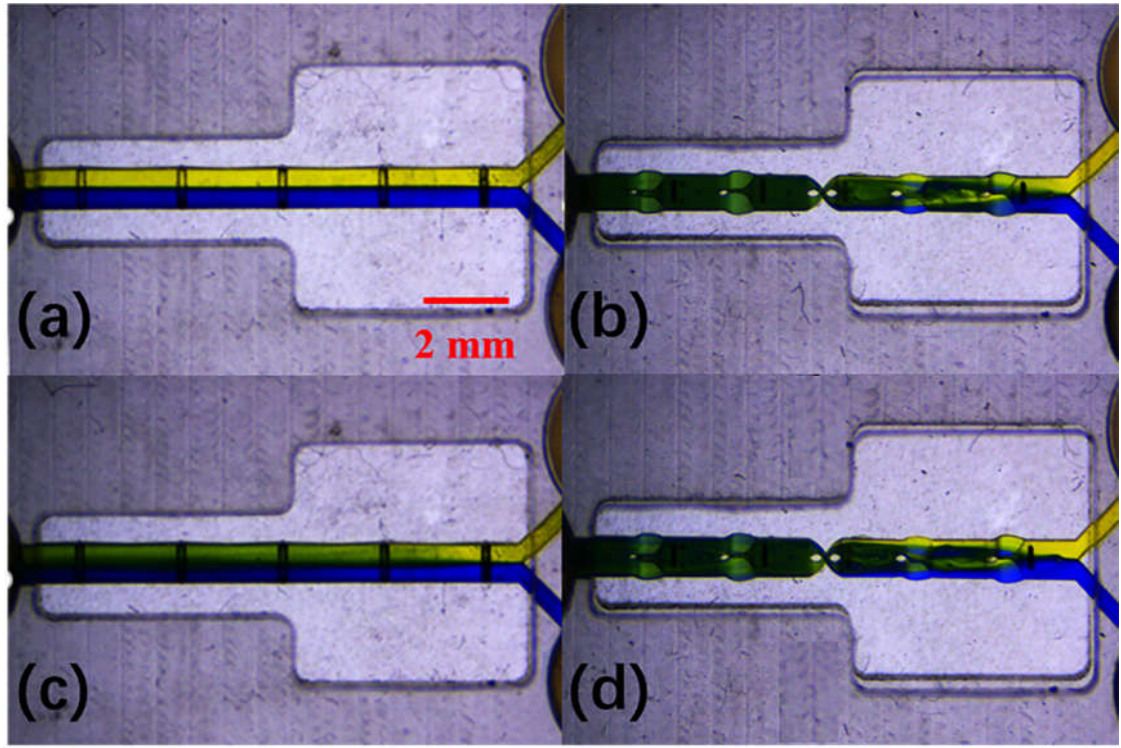

**Figure 5** Microscope images of mixing of blue and yellow ink solutions in (a) 1D straight channel and (b) 3D micromixer at a flow rate of 10 cm s$^{-1}$. For comparison, the same mixing processes in (c) 1D straight channel and (d) 3D micromixer are recorded at a higher flow rate of 30 cm s$^{-1}$. Scale bar in (a)-(d): 2 mm.